# Say it or AI it: Evaluating Hands-Free Text Correction in Virtual Reality

Ziming Li
The Hong Kong University of Science
and Technology (Guanzhou)
Guanzhou, China
zli578@connect.hkust-gz.edu.cn

Joffrey Guilmet
EsieaLab
Laval, France
joffrey.guilmet@esiea.fr

Suzanne Sorli
EsieaLab
Laval, France
suzanne.sorli@esiea.fr

Hai-Ning Liang[*]
The Hong Kong University of Science
and Technology (Guanzhou)
Guanzhou, China
hainingliang@hkust-gz.edu.cn

Diego Monteiro[*]
EsieaLab
Laval, France
diego.vilelamonteiro@esiea.fr

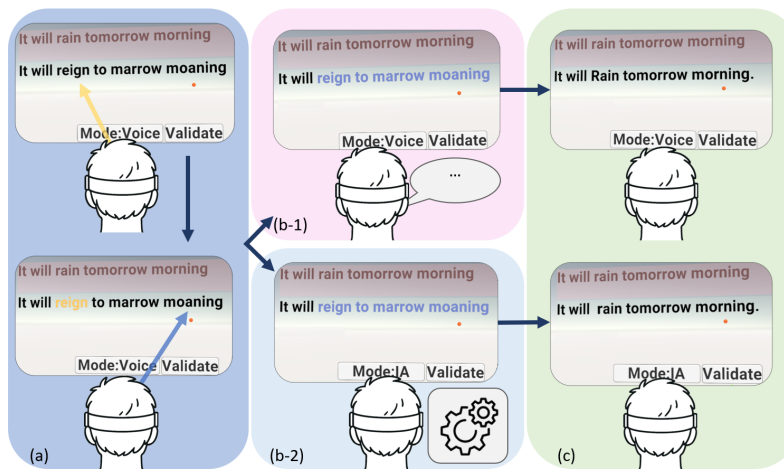

Figure 1: Illustration of our keyboard-free correction approach in a VR context. The user begins by selecting the first and last words of the incorrect segment. Depending on the selected mode via the toggle, the user can either (a) speak to correct the segment using voice input (b-1) or (b) rely on the AI to suggest a context-aware correction (b-2). For single-word selection, an initial dwell highlights the word in yellow, followed by a second dwell turning it blue to confirm selection. For multi-word selection, the user dwells on the first word, then selects the last word, after which the entire segment turns blue. Finally, the sentence is updated with the corrected version according to the chosen method (c).

## Abstract

Text entry in Virtual Reality (VR) is challenging, even when accounting for the use of controllers. Prior work has tackled this challenge head-on, improving the efficiency of input methods. These techniques have the advantage of allowing for relatively straightforward text correction. However, text correction without the use of controllers is a topic that has not received the same amount of attention, even though it can be desirable in several scenarios, and can even be the source of frustration. Large language models have been adopted and evaluated as a corrective methodology, given their high power for predictions. Nevertheless, their predictions are not always correct, which can lead to lower usability. In this paper, we investigate whether, for text correction in VR that is hands-free, the use of AI could surpass in terms of usability and efficiency. We observed better usability for AI text correction when compared to voice input.

[*]Corresponding Author

**ACM Reference Format:**
Ziming Li, Joffrey Guilmet, Suzanne Sorli, Hai-Ning Liang, and Diego Monteiro. 2025. Say it or AI it: Evaluating Hands-Free Text Correction in Virtual Reality. In . ACM, New York, NY, USA, 8 pages.

## 1 Introduction

Text entry in Virtual Reality (VR) environments represents an important yet challenging aspect, especially when considering the absence of traditional input devices such as keyboards or controllers (i.e., hands-free interaction) [Yu et al. 2018a]. To improve efficiency, various techniques utilizing controllers [Boletsis and Kongsvik 2019; Yu et al. 2018a], gestures [Spiess et al. 2022], and gaze-based [Zhao et al. 2023] interaction methods have been developed [Dube and Arif 2019a]. However, the reliance of these methods on physical input devices can limit accessibility in certain VR scenarios. For this reason, achieving efficient and accurate text entry remains an ongoing research concern, prompting extensive research into new techniques and input mechanisms [Giovannelli et al. 2022; Wan et al. 2024a,b]. Among these, hands-free text input methods have emerged as a promising direction to address issues related to accessibility and usability challenges, with positive results [Wan et al. 2024b].

Despite progress in developing effective text entry techniques, there is limited research specifically targeting methods for correcting erroneous input within VR settings. Error correction is particularly crucial since misinterpretations by voice or gesture recognition systems can lead to user frustration and reduced efficiency. Existing studies have examined different approaches, including retyping mistaken words, caret navigation, and multimodal combinations of speech and keyboard inputs [Singh and Singh 2018]. Yet, the integration of AI, particularly Large Language Models (LLMs), presents a promising avenue that has not been thoroughly evaluated in the context of text correction for VR systems. This gap underscores the need for an in-depth investigation into how AI-driven solutions can complement or enhance existing correction mechanisms.

In this paper, we explore two hands-free approaches to text correction for VR environments: one utilizing an LLM to propose context-aware corrections and another employing voice recognition for independent word modifications. Our study aims to evaluate the effectiveness of these methods using both subjective and objective usability measures. The findings indicate that the efficiency of each method varies depending on the frequency and complexity of required corrections. Notably, the combination of both techniques emerges as an optimal solution for most practical scenarios, highlighting the complementary strengths of AI and voice input in enhancing user experience in VR.

This work contributes to the broader understanding of text correction in VR by addressing the nuanced interplay between automated AI predictions and user-driven voice inputs. By focusing on hands-free interactions, our work aligns with the growing demand for more accessible, practical, and inclusive VR experiences [Creed et al. 2024], offering insights that could inform future developments in intelligent, multimodal correction systems. In short, this study underscores the potential of hybrid approaches to overcome existing limitations and pave the way for more robust and user-friendly VR interfaces.

## 2 Related Work

### 2.1 Text Input in Head Mounted Displays

Text input in VR head-mounted displays (HMDs) has been reasonably well-studied. Works such as Yu et al.[Yu et al. 2017] have compared the use of Tap, Dwell, or Gesture as feasible ways to input Text in VR. Others, such as PizzaText [Yu et al. 2018b], have investigated efficient ways to perform fast text input in VR, using controllers and non-standard keyboard layouts. Moreover, we can find several other examples that use head and eye-tracking as methods for input using controllers, keyboards, or virtual keyboards [Dube and Arif 2019b].

Recently, Chen et al. [Chen et al. 2024] proposed using LLMs as a method for predictive text input, leveraging their context awareness. They proposed it both as a way to advance what the user would say and as a way to reduce typing, as the words could be introduced using a simplified spelling, and the LLM would replace the tokens with proper words. Their results yielded 15.50 to 26.65 Words Per Minute (WPM) and, as far as the authors' awareness, this is the first work presenting the use of LLM for such a goal in VR.

Speech-to-text literature is quite ample [Reddy et al. 2023], and it has even identified that people with accents struggle when using shorter phrases [Cumbal et al. 2024]. However, a systematic literature review on Virtual Reality Interactions [Monteiro et al. 2021] identified fewer studies using hands-free symbolic input in VR. Among them, one of the most cited is [Pick et al. 2016], which makes use of speech-to-text but is also multimodal. Other versions found in the literature review were more concerned with other interactions rather than text input. For instance, Hepperle et al. [Hepperle et al. 2019] investigated voice for object interaction, leaving speech-to-text input a somewhat underexplored area in VR.

### 2.2 Text Selection

In immersive environments, selecting text is not limited by the absence of input devices, but by the challenge of designing methods that are fast, accurate, and minimally disruptive [Song et al. 2024]. [Xu et al. 2022] evaluated several techniques for text selection in VR, including raycasting, direct touch, and indirect touch using motion controllers. They found that raycasting was generally faster, especially for longer words, while direct touch offered higher precision. In parallel, other research has explored hands-free approaches that rely on gaze, head movement, or speech. For example, [Meng et al. 2022] compared head gaze, eye gaze, and speech-based selection for both character-level and word-level tasks. Their results showed that eye gaze was the fastest for selecting single characters but required high precision and could lead to fatigue, whereas head gaze was more stable but slower. Although focused on augmented reality, [Liu et al. 2023] provides insights relevant to VR by comparing direct and indirect text selection techniques using gaze and gesture. Their findings showed that direct selection led to better performance and lower mental workload, especially in dense text layouts, suggesting that precision remains a central challenge across immersive environments. Additional insights come from AR-based studies. [Shi et al. 2023] compared unimodal and multimodal region selection techniques, showing that combining gaze with hand input improved both speed and precision. [Lystbæk et al. 2022] further highlighted how gaze-only selection is highly sensitive to calibration and target size, reinforcing the difficulty of achieving reliable precision with gaze-driven methods alone.

### 2.3 Text Correction

Text correction approaches range from typing-based methods to multimodal techniques. Cui et al. [Cui et al. 2020] introduced word retyping with automatic inference for replacement. In immersive environments, caret navigation techniques have been explored: Hu et al. [Hu et al. 2022] compared AR caret placement methods (touch, raycasting, gaze), while Li et al. [Li et al. 2021] found continuous caret movement with word-level backspacing most effective in VR correction tasks.

Multimodal approaches combine different input modalities. Eye-SayCorrect [Zhao et al. 2022] uses gaze for word selection and speech for correction on mobile devices. Other work explored speech-keyboard combinations [Kristensson and Vertanen 2011] and speech-only pipelines [McNair and Waibel 1994; Vertanen and Kristensson 2009, 2010], though these struggle with inference failures and selection efficacy. Adhikary et al. [Adhikary and Vertanen 2021] combined speech with midair hand tracking in VR, using BERT for context-aware correction suggestions, achieving 28 WPM with 0.5% error rate.

Recent comparative studies provide systematic evaluations. Du et al. [Du et al. 2022] analyzed human editing patterns across large datasets, proposing taxonomies of edit types. Dudley et al. [Dudley et al. 2024] compared five VR correction techniques (raycasting, touch methods, gaze, speech), finding gaze fast but inaccurate, with users preferring ray-based and direct-touch methods for their precision-efficiency balance.

While various text correction approaches exist, none have evaluated LLM-powered correction combined with gaze and voice specifically for VR text replacement—existing works differ in device platform or research scope.

## 3 System Design and Implementation

We designed and implemented a hands-free interaction system based on the VR platform. This system was designed for assessing the performance of various sentence correction methods. It targets hands-free scenarios where users must correct the errors generated from automatic speech recognition, which may stem from technological inaccuracies or the user's accent. The system integrates two principal correction methods: one based on direct voice input from the user, and another based on the LLM for automated correction. To systematically compare these methods in user experiments, the system set up three correction modes: Voice-only, AI-only, and a Toggle mode, which permits users to alternate between the two methods at will.

The system's interaction process is divided into two stages: selection and correction. During the selection stage, the user utilizes gaze-based interaction to mark a "Correction Range" within the initial sentence. To streamline the interaction and mitigate cognitive load, the system transitions seamlessly to the correction stage after the "Correction Range" is selected. In this stage, based on the active experimental correction mode, the user either inputs a new voice sentence or triggers the LLM to perform an automatic correction. For the LLM-based correction method (AI-method), a specific prompt is sent to Gemma 3 27B Model, stating: "***Initial Sentence** is a sentence derived from speech input and currently contains errors due to accent and recognition issues. Please revise the **Correction Range** to correct the sentence and return the complete, accurate sentence.*" A detailed illustration of the entire system workflow is provided in Figure 2.

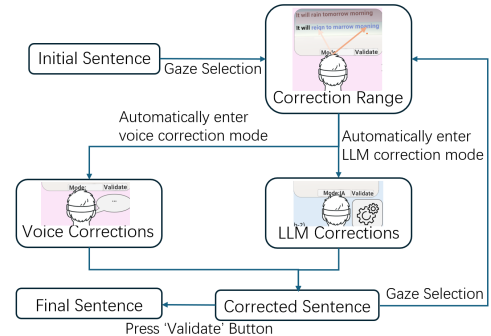

**Figure 2: Workflow of the sentence correction platform.**

This system was set up on an Intel Core i7-13620H processor PC with a dedicated NVIDIA GeForce RTX 4070 graphics card. The program was developed using C#.NET and was run in the Unity3D platform, and we used the Meta Quest 3s as the display equipment. The speech-to-text functionality was implemented using the Meta Voice SDK within the VR environment, with the language set to English. The LLM was executed locally using the Ollama runtime environment, running the Gemma 3 27B Model, on CPU. The server was an Intel Core i5 processor PC with a dedicated NVIDIA GeForce RTX 3080 graphics card and 192 GB of RAM.

## 4 Study 1: Single-Word Correction Task

### 4.1 Experiment Design

Study 1 aims to investigate user performance in the simplified scenario of correcting single-word errors in voice input. We compared all three correction modes: (Condition A) Voice-mode, (Condition B) AI-mode, and (Condition C) Toggle-mode. A within-subjects design was employed, where each participant experienced all three modes and completed a series of sentence correction tasks. The experimental materials comprised 30 sentences, each containing a single error. These sentences were generated by expanding 10 real-world speech recognition error samples using the GPT-4o model. All participants corrected the same set of sentences (10 for each mode), but the presentation order of the sentences was randomized. Furthermore, the sequence of the three correction modes was counterbalanced to mitigate potential ordering effects.

Based on the literature, we have proposed the following hypothesis: for single-word correction tasks, the AI-based correction method will outperform voice input in terms of **H1** efficiency, **H2** accuracy, and **H3** perceived ease of use.

## 4.2 Experiment Procedure

Before the experiment, participants received a thorough briefing. They were informed of their right to withdraw at any point without reason and to retract their data. We also highlighted the potential for cybersickness when using the VR Head-Mounted Display (VR HMD). Upon giving informed consent, each participant filled out a pre-study questionnaire covering demographic information. A training session was held to ensure all participants reached a consistent skill level on our sentence correction system and the three distinct correction methods.

The formal experiment utilized a Latin Square design to counterbalance the order of conditions and minimize learning effects. Participants were exposed to all three correction methods, using each to fix 10 sentences with single-word errors. The system automatically logged performance data for each trial for later data analysis. After the tasks, a post-study questionnaire was administered to collect subjective feedback on the system. Study 1 took approximately 20 to 30 minutes to complete.

## 4.3 Participants

Sixteen participants (3 females, 13 males) aged between 18 and 50 (M = 25.1, SD = 8.55) were recruited from a local university to participate in this study. According to our pre-experiment demographic questionnaire, 13 participants had prior experience with VR beyond brief demos (i.e., more than 1 hour of total use). Additionally, 9 participants reported intermediate or advanced experience with voice input technology.

## 4.4 Analysis and Results

We collected 480 complete behavioral data sets (16 participants × 3 conditions × 10 tasks) and 96 subjective data points. Performance metrics included task completion time, number of corrections, Word Error Rate (WER), and Semantic Error Rate (SER). SER was computed using cosine similarity between target and final sentences encoded with the *all-MiniLM-L6-v2* model. We used Repeated Measures ANOVA (RM-ANOVA) with Bonferroni correction for objective data and Friedman tests with Wilcoxon signed-rank tests for subjective measures (NASA-TLX and SUS).

*4.4.1 Completion Time and Correction Times.* Task completion time showed significant differences ($F(2, 30) = 14.34$, $p < .001$, $\eta_p^2 = .32$). AI-mode was fastest ($M = 17.87s$), followed by Toggle-mode ($M = 24.29s$) and Voice-mode ($M = 34.86s$). All pairwise comparisons were significant. For correction frequency ($F(2, 30) = 12.52$, $p = .002$, $\eta_p^2 = .39$), Voice-mode required significantly more corrections ($M = 2.49$) than both AI-mode ($M = 1.29$) and Toggle-mode ($M = 1.59$), supporting **H1**. The above results are shown in Figure 3, while informing **H1**.

*4.4.2 Error Rates.* Both semantic ($F(2, 30) = 6.42$, $p = .005$) and word error rates ($F(2, 30) = 4.26$, $p = .023$) showed significant effects. Toggle-mode achieved the lowest error rates (SER: $M = 0.05$; WER: $M = 0.05$), significantly outperforming Voice-mode (SER: $M = 0.13$; WER: $M = 0.13$). The above results are shown in Figure 4, while informing **H2**.

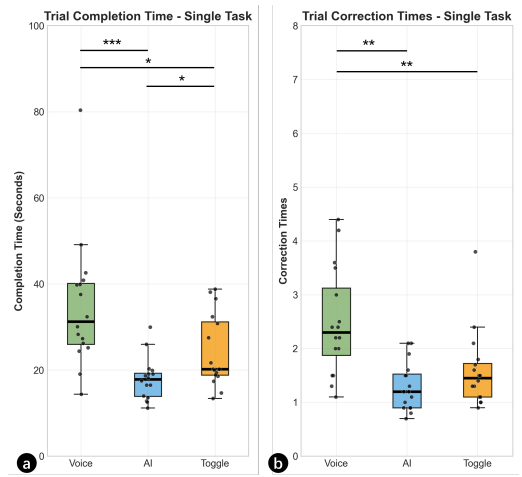

Figure 3: (a) Mean Task Completion Time (seconds) and (b) Mean Task Correction Times (seconds) per trial for each correction mode in Single Word variation.

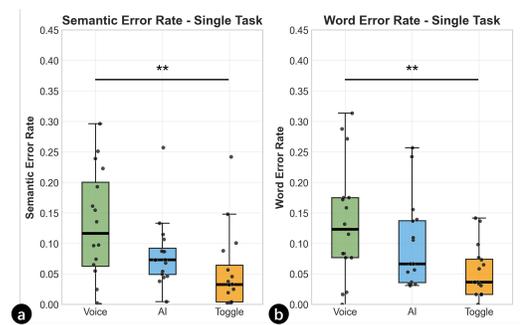

Figure 4: (a) Mean Task Semantic Error Rate (normalized) and (b) Mean Task World Error Rate (normalized) per trial for each correction mode in Single Word variation.

*4.4.3 NASA-TLX and SUS.* NASA-TLX revealed significant differences in Mental Demand, Effort, Frustration, and Performance. Voice-mode was rated significantly more demanding and frustrating than AI-mode and Toggle-mode. SUS scores showed no significant differences. The NASA-TLX results are shown in Figure 5. The above results inform **H3**.

*4.4.4 User Behavior in Toggle Correction Mode.* Users showed strong AI preference (74.4% AI-only trials vs. 7.5% Voice-only), but 93.8% used both methods strategically. Users exhibited "failure tolerance," switching from Voice to AI after 2.06 failures but from AI to Voice after only 1.31 failures, indicating higher expectations for AI performance.

## 4.5 Discussion

Study 1 evaluated three correction modes for single-word errors in hands-free scenarios. The findings largely support our hypotheses, confirming AI-assisted correction advantages for simple tasks while revealing strategic user behavior.

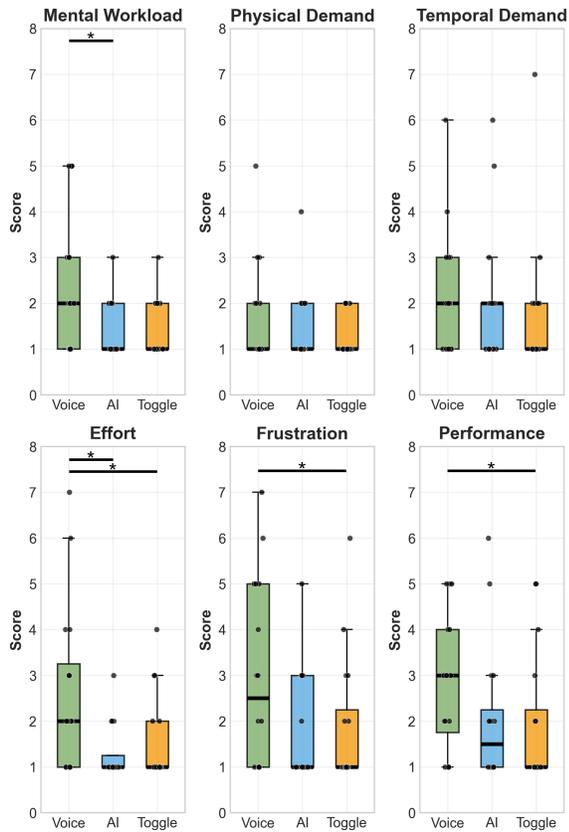

Figure 5: NASA-TLX workload ratings (Mean ± SD) for the single-word input task. Lower scores indicate lower perceived workload, except for Performance, where higher scores indicate worse self-rated performance.

As predicted (**H1**, **H3**), AI-mode excelled in completion time and perceived workload through direct manipulation without re-input, avoiding Voice-mode's temporal costs and "cascading errors" risk. However, **H2** was not supported—Toggle-mode achieved the highest accuracy by providing user control. Behavioral analysis revealed participants adopted an "AI-first" strategy (74.4% of trials) but strategically switched to voice when AI suggestions failed.

Study 1 concludes that for single-word corrections, AI-assisted methods are central to efficiency, but hybrid interfaces providing user control optimize accuracy. However, this study's limitation is its focus on simple tasks. Real-world speech recognition failures involve complex scenarios—multiple incorrect words, grammatical mistakes, or sentence-level reformulations. Performance differences between modes might change under increased complexity, motivating Study 2's investigation of complex correction tasks.

# 5 Study 2: Multi-Word Correction Task
## 5.1 Study Design
In Study 2, we aimed to evaluate and compare the performance of the three revision modes in more complex hands-free text correction scenarios. We followed the same experimental procedure and used the same metrics as in Study 1, and invited the same 16 participants to take part. Different from Study 1, where each sentence contained only a single error, Study 2 increased the task difficulty by including two to four errors per sentence. The experimental materials consisted of 30 sentences. Among these, 10 were collected from real-world contexts, while the remaining 20 were generated by expanding upon the original sentences using OpenAI's GPT-4o model. The entire experiment for Study 2 lasted approximately 30 to 40 minutes.

As a further step from Study 1, we have proposed the following hypothesis: for multiple-word correction tasks, the AI-based correction method will outperform voice input in terms of **H4** efficiency, **H5** accuracy, and **H6** perceived ease of use.

## 5.2 Analysis and Results

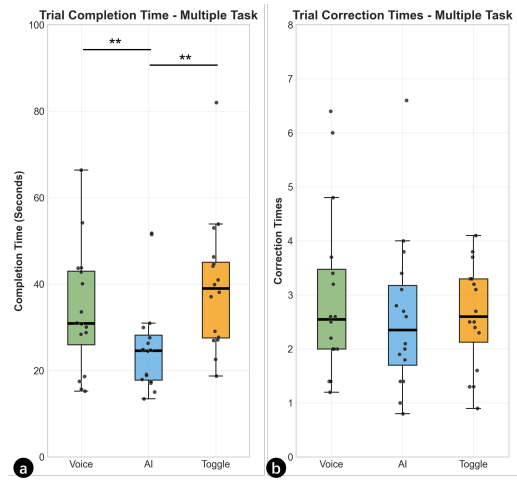

Figure 6: (a) Mean Task Completion Time (seconds) and (b) Mean Task Correction Times (seconds) per trial for each correction mode in Multiple Words variation.

*5.2.1 Completion Time and Correction Times.* RM-ANOVA revealed significant differences in task completion time ($F(2, 30) = 10.87$, $p < .001$, $\eta_p^2 = .15$). AI-mode was fastest ($M = 25.71s$), significantly outperforming Voice-mode ($M = 33.80s$; $p < .01$) and Toggle-mode ($M = 39.53s$; $p < .01$). No significant differences were found between Voice-mode and Toggle-mode. For correction frequency, no significant main effect was observed, though Voice-mode showed the highest corrections ($M = 2.96$), followed by Toggle-mode ($M = 2.63$) and AI-mode ($M = 2.59$). The findings are illustrated in Figure 6 and inform **H4**.

*5.2.2 Error Rates.* Both semantic ($F(2, 30) = 33.95$, $p < .001$, $\eta_p^2 = .69$) and word error rates ($F(2, 30) = 31.93$, $p < .001$, $\eta_p^2 = .54$) showed significant effects. Contrary to Study 1, AI-mode produced the highest error rates (SER: $M = 0.32$; WER: $M = 0.41$), followed by Voice-mode (SER: $M = 0.12$; WER: $M = 0.17$), while Toggle-mode achieved the lowest errors (SER: $M = 0.06$; WER: $M = 0.12$). All pairwise comparisons between AI-mode and other modes were significant. The above results are shown in Figure 7 and contribute to the interpretation of **H5**.

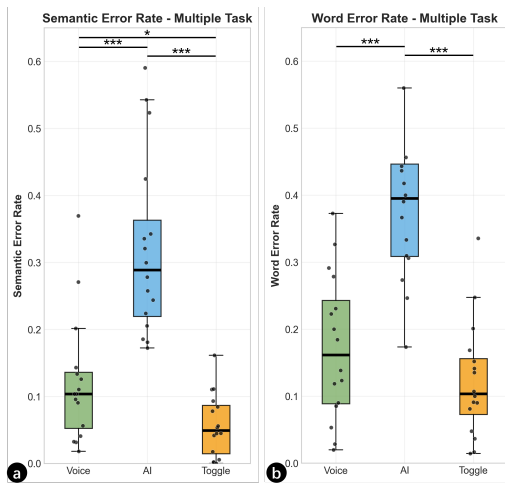

Figure 7: (a) Mean Task Semantic Error Rate (normalized) and (b) Mean Task World Error Rate (normalized) per trial for each correction mode in Multiple Words variation.

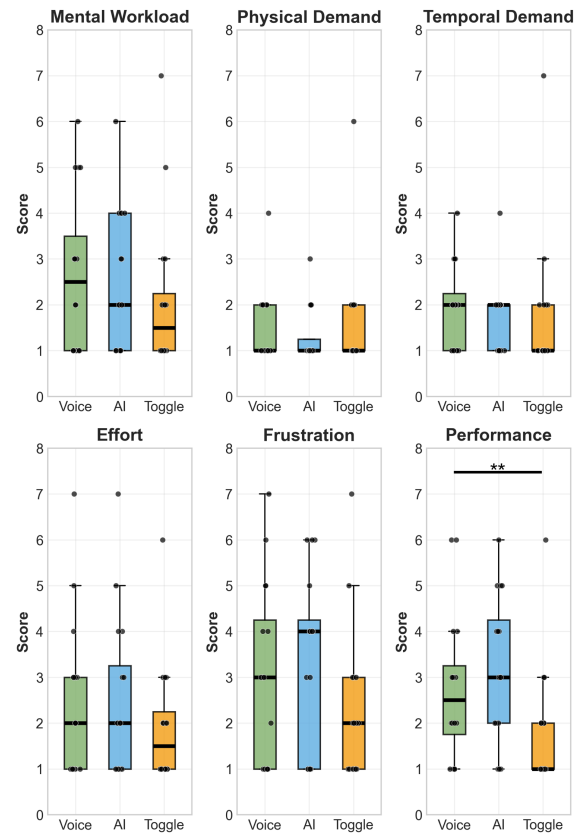

Figure 8: NASA-TLX workload ratings (Mean ± SD) for the multiple-words input task. Lower scores indicate lower perceived workload, except for Performance, where higher scores indicate worse self-rated performance.

*5.2.3 NASA-TLX and SUS.* Friedman tests revealed significant differences in NASA-TLX Performance subscale ($\chi^2(2) = 9.722$, $p = .008$), with Voice-mode rated significantly worse than Toggle-mode. SUS scores showed significant differences ($\chi^2(2) = 12.808$, $p = .002$), with Toggle-mode rated significantly higher in usability than Voice-mode. The NASA-TLX and SUS results are shown in Figure 8. The above findings inform **H6**.

*5.2.4 User Behavior in Toggle Correction Mode.* Users showed strong AI preference (61.9% AI-only trials vs. 4.4% Voice-only), with 93.8% strategically using both modes. Similar to Study 1, users exhibited "failure tolerance," switching from Voice to AI after 2.11 failures and from AI to Voice after 2.03 failures, showing nearly equal patience for both modes in complex tasks.

## 5.3 Discussion

Study 2 investigated how increased task complexity affects the performance of the three interaction modes. The results reveal that task complexity significantly influenced mode performance, particularly AI-mode accuracy.

Regarding efficiency and accuracy (**H4**, **H5**), AI-mode remained fastest for task completion, retaining its speed advantage through direct selection rather than re-input. However, our accuracy hypothesis was refuted—AI-mode yielded significantly higher error rates than Voice-mode and Toggle-mode in multi-word corrections. Complex tasks demand greater contextual understanding from AI, while Voice-mode's high expressive power enabled precise multi-error corrections.

For subjective experience (**H6**), unlike Study 1, no significant cognitive load differences emerged across most dimensions. When tasks are sufficiently difficult, all modes pose comparable cognitive challenges. Evaluating AI suggestions becomes as complex as voice input corrections. Toggle-mode's top SUS score aligns with its best accuracy, indicating users prefer effectiveness and controllability for challenging tasks.

User behavior in Toggle-mode also shifted with complexity. AI failure tolerance increased from 1.31 attempts (Study 1) to 2.03, suggesting users' mental model adaptation—they gave AI more opportunities when facing partially correctable complex errors. However, this tolerance remained lower than for Voice-mode (2.11 attempts), indicating a consistent preference for more controllable methods.

In summary, Study 2 demonstrates that in complex correction scenarios, speed advantages yield to accuracy and user control importance. These findings contrast with Study 1 and provide crucial insights for understanding interaction mode selection under varying task complexity.

## 6 General Discussion
### 6.1 Key Findings
This research explores user behavior and preferences in VR-based hands-free text correction. We conducted two experiments of increasing difficulty. We systematically analyzed three interaction modes across efficiency, accuracy, and user experience dimensions.

First, AI-mode shows efficiency-accuracy trade-offs as task complexity increases. AI-mode consistently achieved the shortest completion times in both simple and complex tasks. Its direct correction without re-input confirmed the speed advantage. However, accuracy decreased sharply with complexity. AI-mode maintained acceptable rates for single-error corrections but performed significantly worse than Voice-mode for multiple, non-contiguous errors. This efficiency-accuracy decoupling highlights current LLMs' dependency on contextual information for complex tasks.

Second, Toggle-mode demonstrated a strong correlation between accuracy and user experience. Toggle-mode empowered users with mode choice and showed exceptional robustness. It consistently achieved the highest correction accuracy across both studies. This proves that users prefer efficient AI tools initially. However, retaining high-control methods as complements is crucial for quality outcomes. Subjective feedback supports this finding. Toggle-mode achieved significantly higher SUS scores in Study 2. Users focus on controllability and reliability for challenging tasks.

Finally, users employed adaptive strategies linked to task difficulty. AI error tolerance varied with complexity. Users showed low tolerance in simple tasks (1.31 failed attempts). They increased patience for complex tasks (2.03 attempts). This nearly matched voice input tolerance (2.11). Users adjust AI performance expectations based on task difficulty. They become more accepting of mistakes for harder problems.

### 6.2 Design Implications for Hands-Free Text-Entry Systems
This research provides key design implications for hands-free text entry and editing systems. Our findings show that the optimal approach is not to choose between a fully user-controlled method and a fully AI-driven one, but to effectively combine them. We argue that future designers should move beyond searching for a single best mode and instead equip users with a mixed-mode toolkit of smart interaction methods.

Our studies revealed that neither the Voice-mode nor the AI-mode correction method does the best in all situations. The AI-mode proved efficient for simple corrections (Study 1) but suffered from a significant decrease in accuracy with complex errors (Study 2). In contrast, the Voice-mode, despite its high controllability, was consistently slow and risked introducing new errors. The success of the Toggle-based mode, which scored highest in accuracy and usability across both studies, was due to its recognition of the complementary strengths of AI and user control, ultimately placing the mode choice decision in the user's hands. An ideal system should, therefore, feature AI assistance as a primary means to boost efficiency, while preserving traditional interaction methods as a reliable fallback for ensuring accuracy and managing complex cases.

## 7 Limitations and Future Work
While this study offers valuable insights into hands-free text correction in VR, several limitations should be considered when interpreting the results. These limitations also point toward important directions for future research.

Our study has four main limitations. First, the participant pool of 16 individuals, though sufficient for a within-subjects design, may not fully represent diverse user behaviors and accents. The simulated phonetic challenges based on non-native accents do not cover the full spectrum of real-world acoustic variations. Second, our offline LLM setup (Gemma 3) was necessary for experimental control but may differ from future cloud-based systems in performance. The LLM received only incorrectly transcribed sentences and brief error descriptions, lacking a broader conversational context that could enhance correction capabilities. Third, the constrained correction tasks (single words or specific phrases within predefined sentences) enabled systematic comparison but do not capture the complexity of real-world VR interactions such as long-form writing or multitasking, leaving the system's scalability to these realistic scenarios unclear. Fourth, our error analysis is reported as aggregate rates. The lack of a fine-grained breakdown by error type or failure cases limits our specific insights into how to further refine the AI and interaction design.

Building upon these limitations, our future research will proceed in three directions. We will refine the user interface by integrating candidate-list approaches, presenting users with ranked correction alternatives rather than single automated corrections, especially beneficial for ambiguous cases. We will enhance LLM performance by expanding contextual inputs beyond current text and scenario descriptions to include chat history, environmental sensor data, and user voice re-input for handling complex errors. To address the lack of detailed failure insights, we will conduct granular error analyses to categorize failure modes and guide targeted model improvements. Finally, we will evaluate our correction interface within ecologically valid VR use cases—voice-based note-taking, collaborative document editing, and synchronous messaging—to test system scalability under practical stressors like cognitive load and divided attention.

## 8 Conclusion
Text correction in virtual reality that is hands-free remains a challenging, yet crucial, aspect of enabling fluid and efficient communication in immersive environments. Our study investigated three approaches for correcting misrecognized input in Voice—mode, AI-mode, and Toggle-mode. Across both single-word and multiple-word correction tasks, our results demonstrate that the hybrid approach is vastly superior.

While voice input provides a familiar and natural modality, its limitations in phonetic disambiguation and syntactic consistency were evident, particularly when no surrounding context was leveraged. LLM-based correction significantly improved performance in constrained single-word scenarios, but its effectiveness dropped sharply in complex multi-word cases, possibly due to inference errors and context degradation. These findings highlight the strength of multimodal flexibility in VR interactions rather than relying solely on automation or natural language inference, allowing users to shift between efficiency and control dynamically.

This work has meaningful implications for accessibility and education in immersive contexts. For users with motor impairments or those unable to use physical controllers, the ability to correct text using gaze and voice offers a vital communication pathway. Furthermore, in educational settings where VR is increasingly used for language learning, writing support, and collaboration, effective correction tools can help learners focus on meaning and fluency rather than being penalized for transient or phonetic errors.

# References

Jiban Adhikary and Keith Vertanen. 2021. Text Entry in Virtual Environments using Speech and a Midair Keyboard. *IEEE Transactions on Visualization and Computer Graphics* 27, 5 (2021), 2648–2658. doi:10.1109/TVCG.2021.3067776

Costas Boletsis and Stian Kongsvik. 2019. Controller-based text-input techniques for virtual reality: An empirical comparison. *International Journal of Virtual Reality* 19, 3 (2019), 2–15.

Liuqing Chen, Yu Cai, Ruyue Wang, Shixian Ding, Yilin Tang, Preben Hansen, and Lingyun Sun. 2024. Supporting Text Entry in Virtual Reality with Large Language Models. 524–534. doi:10.1109/VR58804.2024.00073

Chris Creed, Maadh Al-Kalbani, Arthur Theil, Sayan Sarcar, and Ian Williams. 2024. Inclusive AR/VR: accessibility barriers for immersive technologies. *Universal Access in the Information Society* 23, 1 (2024), 59–73.

Wenzhe Cui, Suwen Zhu, Mingrui Ray Zhang, H Andrew Schwartz, Jacob O Wobbrock, and Xiaojun Bi. 2020. Justcorrect: Intelligent post hoc text correction techniques on smartphones. In *Proceedings of the 33rd Annual ACM Symposium on User Interface Software and Technology*. 487–499.

Ronald Cumbal, Birger Moell, José Lopes, and Olof Engwall. 2024. You don't understand me!: Comparing ASR results for L1 and L2 speakers of Swedish. doi:10.48550/arXiv.2405.13379

Wanyu Du, Vipul Raheja, Dhruv Kumar, Zae Myung Kim, Melissa Lopez, and Dongyeop Kang. 2022. Understanding iterative revision from human-written text. *arXiv preprint arXiv:2203.03802* (2022).

Tafadzwa Joseph Dube and Ahmed Sabbir Arif. 2019a. Text entry in virtual reality: A comprehensive review of the literature. In *Human-Computer Interaction. Recognition and Interaction Technologies: Thematic Area, HCI 2019, Held as Part of the 21st HCI International Conference, HCII 2019, Orlando, FL, USA, July 26–31, 2019, Proceedings, Part II 21*. Springer, 419–437.

Tafadzwa Joseph Dube and Ahmed Sabbir Arif. 2019b. Text Entry in Virtual Reality: A Comprehensive Review of the Literature. In *Human-Computer Interaction. Recognition and Interaction Technologies*, Masaaki Kurosu (Ed.). Springer International Publishing, Cham, 419–437.

John J. Dudley, Amy Karlson, Kashyap Todi, Hrvoje Benko, Matt Longest, Robert Wang, and Per Ola Kristensson. 2024. Efficient Mid-Air Text Input Correction in Virtual Reality. In *2024 IEEE International Symposium on Mixed and Augmented Reality (ISMAR)*. 893–902. doi:10.1109/ISMAR62088.2024.00105

Alexander Giovannelli, Lee Lisle, and Doug A Bowman. 2022. Exploring the impact of visual information on intermittent typing in virtual reality. In *2022 IEEE international Symposium on Mixed and augmented reality (ISMAR)*. IEEE, 8–17.

Daniel Hepperle, Yannick Weiß, Andreas Siess, and Matthias Wölfel. 2019. 2D, 3D or speech? A case study on which user interface is preferable for what kind of object interaction in immersive virtual reality. *Computers & Graphics* 82 (2019), 321–331. doi:10.1016/j.cag.2019.06.003

Jinghui Hu, John J Dudley, and Per Ola Kristensson. 2022. An evaluation of caret navigation methods for text editing in augmented reality. In *2022 IEEE International Symposium on Mixed and Augmented Reality Adjunct (ISMAR-Adjunct)*. IEEE, 640–645.

Per Ola Kristensson and Keith Vertanen. 2011. Asynchronous Multimodal Text Entry Using Speech and Gesture Keyboards. In *Proceedings of the Twelfth Annual Conference of the International Speech Communication Association (INTERSPEECH)*.

Yang Li, Sayan Sarcar, Yilin Zheng, and Xiangshi Ren. 2021. Exploring text revision with backspace and caret in virtual reality. In *Proceedings of the 2021 CHI conference on human factors in computing systems*. 1–12.

Xinyi Liu, Xuanru Meng, Becky Spittle, Wenge Xu, BoYu Gao, and Hai-Ning Liang. 2023. Exploring Text Selection in Augmented Reality Systems. In *Proceedings of the 18th ACM SIGGRAPH International Conference on Virtual-Reality Continuum and Its Applications in Industry* (Guangzhou, China) *(VRCAI '22)*. Association for Computing Machinery, New York, NY, USA, Article 35, 8 pages. doi:10.1145/3574131.3574459

Mathias N. Lystbæk, Ken Pfeuffer, Jens Emil Sloth Grønbæk, and Hans Gellersen. 2022. Exploring Gaze for Assisting Freehand Selection-based Text Entry in AR. *Proc. ACM Hum.-Comput. Interact.* 6, ETRA, Article 141 (May 2022), 16 pages. doi:10.1145/3530882

A. E. McNair and A. Waibel. 1994. Improving recognizer acceptance through robust, natural speech repair. In *Proceedings of the 3rd International Conference on Spoken Language Processing (ICSLP)*. ISCA.

Xuanru Meng, Wenge Xu, and Hai-Ning Liang. 2022. An Exploration of Hands-free Text Selection for Virtual Reality Head-Mounted Displays . In *2022 IEEE International Symposium on Mixed and Augmented Reality (ISMAR)*. IEEE Computer Society, Los Alamitos, CA, USA, 74–81. doi:10.1109/ISMAR55827.2022.00021

Pedro Monteiro, Guilherme Gonçalves, Hugo Coelho, Miguel Melo, and Maximino Bessa. 2021. Hands-free interaction in immersive virtual reality: A systematic review. *IEEE Transactions on Visualization and Computer Graphics* 27, 5 (2021), 2702–2713. doi:10.1109/TVCG.2021.3067687

Sebastian Pick, Andrew S. Puika, and Torsten W. Kuhlen. 2016. SWIFTER: Design and evaluation of a speech-based text input metaphor for immersive virtual environments. In *2016 IEEE Symposium on 3D User Interfaces (3DUI)*. 109–112. doi:10.1109/3DUI.2016.7460039

V. Madhusudhana Reddy, T. Vaishnavi, and K. Pavan Kumar. 2023. Speech-to-Text and Text-to-Speech Recognition Using Deep Learning. In *2023 2nd International Conference on Edge Computing and Applications (ICECAA)*. 657–666. doi:10.1109/ICECAA58104.2023.10212222

Rongkai Shi, Yushi Wei, Xueying Qin, Pan Hui, and Hai-Ning Liang. 2023. Exploring Gaze-assisted and Hand-based Region Selection in Augmented Reality. *Proc. ACM Hum.-Comput. Interact.* 7, ETRA, Article 160 (May 2023), 19 pages. doi:10.1145/3591129

Shashank Singh and Shailendra Singh. 2018. Review of real-word error detection and correction methods in text documents. In *2018 second international conference on electronics, communication and aerospace technology (ICECA)*. IEEE, 1076–1081.

Jianbin Song, Rongkai Shi, Yue Li, BoYu Gao, and Hai-Ning Liang. 2024. Exploring Controller-based Techniques for Precise and Rapid Text Selection in Virtual Reality. In *2024 IEEE Conference Virtual Reality and 3D User Interfaces (VR)*. 244–253. doi:10.1109/VR58804.2024.00047

Florian Spiess, Philipp Weber, and Heiko Schuldt. 2022. Direct interaction word-gesture text input in virtual reality. In *2022 IEEE International Conference on Artificial Intelligence and Virtual Reality (AIVR)*. IEEE, 140–143.

Keith Vertanen and Per Ola Kristensson. 2009. Automatic selection of recognition errors by respeaking the intended text. In *2009 IEEE Workshop on Automatic Speech Recognition & Understanding (ASRU)*. 130–135. doi:10.1109/ASRU.2009.5373347

Keith Vertanen and Per Ola Kristensson. 2010. Getting it right the second time: Recognition of spoken corrections. In *2010 IEEE Spoken Language Technology Workshop (SLT)*. 289–294. doi:10.1109/SLT.2010.5700866

Tingjie Wan, Yushi Wei, Rongkai Shi, Junxiao Shen, Per Ola Kristensson, Katie Atkinson, and Hai-Ning Liang. 2024a. Design and evaluation of controller-based raycasting methods for efficient alphanumeric and special character entry in virtual reality. *IEEE Transactions on Visualization and Computer Graphics* 30, 9 (2024), 6493–6506.

Tingjie Wan, Liangyuting Zhang, Hongyu Yang, Pourang Irani, Lingyun Yu, and Hai-Ning Liang. 2024b. Exploration of foot-based text entry techniques for virtual reality environments. In *Proceedings of the 2024 CHI Conference on Human Factors in Computing Systems*. 1–17.

Wenge Xu, Xuanru Meng, Kangyou Yu, Sayan Sarcar, and Hai-Ning Liang. 2022. Evaluation of Text Selection Techniques in Virtual Reality Head-Mounted Displays . In *2022 IEEE International Symposium on Mixed and Augmented Reality (ISMAR)*. IEEE Computer Society, Los Alamitos, CA, USA, 131–140. doi:10.1109/ISMAR55827.2022.00027

Difeng Yu, Kaixuan Fan, Heng Zhang, Diego Monteiro, Wenge Xu, and Hai-Ning Liang. 2018a. PizzaText: Text entry for virtual reality systems using dual thumbsticks. *IEEE transactions on visualization and computer graphics* 24, 11 (2018), 2927–2935.

Difeng Yu, Kaixuan Fan, Heng Zhang, Diego Monteiro, Wenge Xu, and Hai-Ning Liang. 2018b. PizzaText: Text Entry for Virtual Reality Systems Using Dual Thumbsticks. *IEEE Transactions on Visualization and Computer Graphics* 24, 11 (2018), 2927–2935. doi:10.1109/TVCG.2018.2868581

Nelson Yu, Deepak Ganesan, Hyunyoung Yeo, and Joseph J. LaViola Jr. 2017. Tap, dwell, and gesture: A quantitative evaluation of microgestures for input on head-mounted displays. In *Proceedings of the 2017 CHI Conference on Human Factors in Computing Systems*. ACM. doi:10.1145/3025453.3025964

Maozheng Zhao, Henry Huang, Zhi Li, Rui Liu, Wenzhe Cui, Kajal Toshniwal, Ananya Goel, Andrew Wang, Xia Zhao, Sina Rashidian, Furqan Baig, Khiem Phi, Shumin Zhai, IV Ramakrishnan, Fusheng Wang, and Xiaojun Bi. 2022. EyeSayCorrect: Eye Gaze and Voice Based Hands-free Text Correction for Mobile Devices. In *Proceedings of the 27th International Conference on Intelligent User Interfaces* (Helsinki, Finland) *(IUI '22)*. Association for Computing Machinery, New York, NY, USA, 470–482. doi:10.1145/3490099.3511103

Maozheng Zhao, Alec M Pierce, Ran Tan, Ting Zhang, Tianyi Wang, Tanya R Jonker, Hrvoje Benko, and Aakar Gupta. 2023. Gaze speedup: Eye gaze assisted gesture typing in virtual reality. In *Proceedings of the 28th International Conference on Intelligent User Interfaces*. 595–606.